\documentclass[12pt]{iopart}

\usepackage{graphicx}
\usepackage{cite}
\begin{document}

\title[Lie symmetry analysis and exact solutions for a variable coefficient Gardner equation ]{Lie symmetry analysis and exact solutions for a variable coefficient Gardner equation arising in arterial mechanics}
%\title{Symmetry Analysis and Some New Exact Solutions For a Variable Coefficient Gardner Equation Arising in Arterial Mechanics}
\renewcommand{\thefootnote}
{\ensuremath{\fnsymbol{footnote}}}
\author{ M~S~Abdel Latif \footnote{permanent address: Engineering Mathematics and Physics
Dept.,
            Faculty of Engineering, Mansoura University, Mansoura,
            Egypt.}}
\address{Department of Applied Mathematics, Astrakhan State University, 20a Tatishchev St., Astrakhan, 414056,
        Russia.}
\ead{m\_gazia@hotmail.com}

%\institute{M.~S.~Abdel Latif \at
 %             Department of Applied Mathematics, Astrakhan State University, 20a Tatishchev St., Astrakhan, 414056,
  %      Russia. \\
   %           \email{m\_gazia@hotmail.com}           \\
 %\and
  %         M.~S.~Abdel Latif \at
%Engineering Mathematics and Physics Dept.,
 %             Faculty of Engineering, Mansoura University, Mansoura, Egypt.
%}
%\address{Department of Applied Mathematics, Astrakhan State University, 20a Tatishchev
%St.\\
%Astrakhan, 414056, Russia \\
 %\email{m\_gazia@hotmail.com}}

\begin{abstract}
In this paper, a variable-coefficient Gardner equation is
considered. By using the classical symmetry analysis method
symmetries for this equation are obtained. Then, the generalized
Jacobi elliptic function expansion method is used to solve the
reduced ODE. Some new exact solutions for the considered PDE are
obtained.
%\keywords{Variable-coefficient Gardner equation; Classical
%Symmetries; Generalized Jacobi elliptic function expansion method,
%Exact solution.}
 %\subclass{ 35C08 \and  34A34 \and  76M60}
\end{abstract}
% 02.30.Ik Integrable systems  02.30.Jr Partial differential equations 05.45.Yv Solitons 04.20.Jb Exact solutions
\maketitle

\section{Introduction}
\label{sec:intro}

The investigation of exact solutions to nonlinear evolution
equations plays an important role in the study of nonlinear physical
phenomena. The wave phenomena are observed in physics, mechanics,
biology, etc. The effort to find these solutions is significant for
the understanding of many physical phenomena, thus they may give
more insight into the physical aspects of the problems.

This paper is devoted to study the solutions of the variable
coefficient Gardner equation which is given by
\begin{equation}\label{eq:MKDV}
    u_t+\mu_1 u u_x+\mu_2u^2 u_x+\mu_3u_{xxx}+h(t)u_x=0,
\end{equation}
where $\mu_1$, $\mu_2$ and $\mu_3$ are constants and $h(t)$ is a
function of $t$. \Eref{eq:MKDV} includes considerably interesting
equations, such as KdV equation, when $\mu_2$ and $h(t)$ are equal
zero, and mKdV equation, when $\mu_1$ and $h(t)$ are equal zero. In
arterial mechanics, some special cases of
\Eref{eq:MKDV} were considered as follows:\\
\begin{enumerate}
  \item The variable
coefficient KdV equation
\begin{equation}\label{eq:vcKDV}
    u_t+\mu_1 u u_x+\mu_3u_{xxx}+h(t)u_x=0,
\end{equation}
was considered
in~\cite{Demiray20091,Demiray20081,Demiray20082,Demiray20072,Demiray20052,Demiray20092,Demiray20071}.
  \item The variable coefficient modified KdV equation
\begin{equation}\label{eq:vcmKDV}
    u_t+\mu_2 u^2 u_x+\mu_3u_{xxx}+h(t)u_x=0,
\end{equation}
was considered in \cite{Demiray20092,Demiray20071}.
  \item The KdV equation
\begin{equation}\label{eq:KDV}
    u_t+\mu_1 u u_x+\mu_3u_{xxx}=0,
\end{equation}
was considered in
~\cite{Demiray20031,Demiray20021,Demiray20001,Demiray19991,Demiray19981,Demiray19982,Demiray19983,Demiray19971,Demiray19961,kudryashov2006,Demiray20041,kudryashov2008}.
  \item The modified KdV equation
\begin{equation}\label{eq:mKDV}
    u_t+\mu_2 u^2 u_x+\mu_3u_{xxx}=0,
\end{equation}
was considered in ~\cite {Demiray20041,kudryashov2008,Demiray20051}.
\end{enumerate}

Special cases of \eref{eq:MKDV} have been studied in ~\cite
{Malfliet2004,Fu2004,Wazwaz2007,Liu2009,bekir2009}. The rest of this
paper is arranged as follows. In section \ref{sec:symmetries}, we
apply the Lie classical symmetry analysis method to \eref{eq:MKDV}.
In Sec. \ref{sec:jacobimethod}, the mathematical framework of the
generalized Jacobi elliptic function expansion method will be
provided. In Sec. \ref{subsec:exactsolutions}, some new exact
solutions of \eref{eq:MKDV} are obtained and application of the
obtained solutions in arterial mechanics will be presented in
section \ref{subsec:Applications}.
%==========================================================================================
%==========================================================================================
%====================              classical symmetries        ============================
%===========================================================================================
%===========================================================================================
\section{Classical symmetries}
\label{sec:symmetries} To apply the Lie classical symmetry analysis
method \cite{olver1993,bluman1989} to \eref{eq:MKDV}, we consider
the one-parameter Lie group of infinitesimal transformations in $x$,
$t$, and $u$ given by
\begin{equation}\label{eq:transformation1}
    x^*=x+\epsilon\xi(x,t,u)+O(\epsilon^2),
\end{equation}
\begin{equation}\label{eq:transformation2}
    t^*=t+\epsilon\tau(x,t,u)+O(\epsilon^2),
\end{equation}
\begin{equation}\label{eq:transformation3}
    u^*=u+\epsilon\eta(x,t,u)+O(\epsilon^2),
\end{equation}
where $\epsilon$ is the group parameter. We require that the set of
solutions of \eref{eq:MKDV} be invariant under this transformation.
This yields an overdetermined system of linear equations for the
infinitesimals $\xi(x, t, u)$, $\tau(x, t, u)$, and $\eta(x, t, u)$.
The associated Lie algebra of infinitesimal symmetries is the set of
infinitesimal generators of the
form\begin{equation}\label{eq:infgenerator}
    v=\xi(x,t,u)\frac{\partial}{\partial x}+\tau(x, t, u)\frac{\partial}{\partial t}+\eta(x, t, u))\frac{\partial}{\partial u}.
\end{equation}
Invariance of \eref{eq:MKDV} under a Lie group of point
transformations with infinitesimal generator \eref{eq:infgenerator}
leads to a set of determining equations. Solving this system we will
obtain:
\begin{equation}\label{eq:tau1}
    \tau=6c_1\mu_2 t+c_2 ,
\end{equation}
\begin{equation}\label{eq:xi1}
    \xi=c1\left(2\mu_2 x-\mu_1^2 t+6\mu_2 th(t)-2\mu_2 \int h(t)dt\right)+c2 h(t)+c3,
\end{equation}
\begin{equation}\label{eq:eta1}
    \eta=-c1(2\mu_2 u+\mu_1),
\end{equation}
where, $c_1$, $c_2$ and $c_3$ are arbitrary constants. The
associated infinitesimal generators are given by:
\begin{equation}\label{eq:infgenerator11}
   \fl v_{1}=\left(2\mu_2 x-\mu_1^2 t+6\mu_2 th(t)-2\mu_2 \int h(t)dt\right)\frac{\partial}{\partial x}+6\mu_2 t\frac{\partial}{\partial t}-(2\mu_2 u+\mu_1)\frac{\partial}{\partial u},
\end{equation}
\begin{equation}\label{eq:infgenerator12}
  \fl  v_{2}= \frac{\partial}{\partial x},
\end{equation}
\begin{equation}\label{eq:infgenerator13}
   \fl v_{3}=h(t)\frac{\partial}{\partial x}+\frac{\partial}{\partial
    t}.
\end{equation}
When considering the infinitesimal generator $v_{1}$, we will obtain
the surface condition
\begin{equation}\label{eq:surface1}
    \left(2\mu_2 x-\mu_1^2 t+6\mu_2 th(t)-2\mu_2 \int h(t)dt\right)\frac{\partial u}{\partial x}+6\mu_2 t\frac{\partial u}{\partial t}=-(2\mu_2 u+\mu_1),
\end{equation}
which when solving we will obtain the similarity
transformation\begin{equation}\label{eq:simtransform1}
    u=\frac{-\mu_1}{2\mu_2}+t^\frac{-1}{3} f(\zeta), \zeta=xt^\frac{-1}{3}+\frac{\mu_1^2}{4\mu_2}t^\frac{2}{3}-t^\frac{-1}{3}\int h(t)dt,
\end{equation}
when substituting from \eref{eq:simtransform1} into \eref{eq:MKDV}
we will obtain the following ordinary differential equation
\begin{equation}\label{eq:ode1}
    - f-\zeta f'+3 \mu_2 f^2f'+3 \mu_3 f'''=0.
\end{equation}
In general, the exact solution for this nonlinear ODE can't be
obtained by using the elementary functions. The series solution of
this equation was obtained in \cite{liu2008}.\\
  When considering the infinitesimal generator
$v_2+\alpha v_3$, $\alpha$ is a constant, we will obtain the surface
condition\begin{equation}\label{eq:surface}
    ( h(t)+\alpha)\frac{\partial u}{\partial x}+\frac{\partial u}{\partial t}=0,
\end{equation}
which when solving we will obtain the similarity
transformation\begin{equation}\label{eq:simtransform}
    u=f(\zeta), \zeta=x-\int h(t)dt-\alpha t,
\end{equation}
when substituting from \eref{eq:simtransform} into \eref{eq:MKDV} we
will obtain the following ordinary differential equation
\begin{equation}\label{eq:ode}
    -\alpha f'+\mu_1 ff'+\mu_2 f^2f'+\mu_3 f'''=0,
\end{equation}
this equation will be solved in the following section.
%==========================================================================================
%==========================================================================================
%==========================            jacobi method           ============================
%===========================================================================================
%===========================================================================================
\section{Generalized
Jacobi elliptic function expansion method} \label{sec:jacobimethod}
For solving \eref{eq:ode} we will use the generalized Jacobi
elliptic function expansion method \cite{zhao2006,wakil2007}. It is
assumed that \eref{eq:ode} has the solutions in the form
\begin{equation}\label{eq:solutionformode}
    u=a_0+\sum_{i=-n}^{n} a_i \phi^i,
\end{equation}
where $a_i$ are constants to be determined later and $\phi$ satisfy
the following elliptic equations
\begin{equation}\label{eq:ellipticeq}
    \phi'^2=r+p\phi^2+q\phi^4,
\end{equation}
furthermore we can get
\begin{equation}\label{eq:ellipticeq1}
    \phi''=p\phi+q\phi^3,
\end{equation}
 where the primes denotes the derivatives with respect to $\zeta$ and $r$,
$p$, $q$ are constants. The solutions of \eref{eq:ellipticeq} are
written in Appendix A \cite{feng2009}. Balancing the highest
derivative term $f'''$ with nonlinear term $ff'$ in \eref{eq:ode}
gives $n=1$, from which we have
\begin{equation}\label{eq:solutionformode1}
    u=a_0+a_1 \phi+a_{-1} \phi^{-1}.
\end{equation}
 Substituting \eref{eq:solutionformode1} along
with \eref{eq:ellipticeq} and \eref{eq:ellipticeq1} into
\eref{eq:ode} and collect all terms with the same powers in
$\phi^k\phi'^l(k=0,1,2,...,l=0,1,...)$  and set the coefficients of
these terms to zero yields a system of nonlinear algebraic equations
for $a_0$, $a_1$ and $a_{-1}$. Solving this system, we find three
sets of solutions.\\
 The first set:
\begin{equation}\label{eq:solution1}
 \fl   a_0=\frac {-\mu_1}{2\mu_2},   a_1=0,      a_{-1}=\pm \sqrt {\frac
    {-6r\mu_3}{\mu_2}},  \alpha=\frac{4p\mu_2 \mu_3-\mu_1^2}{4
    \mu_2}.
\end{equation}
 The second set:
\begin{equation}\label{eq:solution2}
\fl    a_0=\frac {-\mu_1}{2\mu_2},   a_1=\pm \sqrt {\frac
{-6q\mu_3}{\mu_2}},
    a_{-1}=0,  \alpha=\frac{4p\mu_2 \mu_3-\mu_1^2}{4
    \mu_2}.
\end{equation}
The third set:
\begin{equation}\label{eq:solution3}
 \fl   a_0=\frac {-\mu_1}{2\mu_2},   a_1=\pm \sqrt {\frac {-6q\mu_3}{\mu_2}},      a_{-1}=\pm \sqrt {\frac
    {-6r\mu_3}{\mu_2}}, \alpha=\frac{-4\mu_2 \mu_3(\pm 6\sqrt
    {rq}-p)-\mu_1^2}{4\mu_2}.
\end{equation}
%==========================================================================================
%==========================================================================================
%============================              exact solutions      ===========================
%===========================================================================================
%===========================================================================================
\section{Exact solutions}\label{subsec:exactsolutions} In this section
we will give some exact solutions for \eref{eq:MKDV}. Substituting from \eref{eq:solution1}, \eref{eq:solution2} and \eref{eq:solution3} into \eref{eq:solutionformode1} and using the table in Appendix A we will obtain the following solutions of \eref{eq:MKDV}:\\
%==========================================================================================
%==========================================================================================
%=======================            Soliton and soliton-like solutions      ===============
%===========================================================================================
%===========================================================================================
Case 1: Soliton and soliton-like solutions
\begin{equation}\label{eq:family1}
    \fl u_1=\frac{-\mu_1}{2\mu_2}\pm \sqrt{\frac{-6\mu_3}{\mu_2}} \mathrm{csch}\left(x-\int h(t)dt-\left(\frac{4\mu_2\mu_3-\mu_1^2}{4\mu_2}\right) t\right).
\end{equation}
\begin{equation}\label{eq:family2}
    \fl u_2=\frac{-\mu_1}{2\mu_2}\pm \sqrt{\frac{6\mu_3}{\mu_2}}  \mathrm{sech}\left(x-\int h(t)dt-\left(\frac{4\mu_2\mu_3-\mu_1^2}{4\mu_2}\right) t\right).
\end{equation}
\begin{equation}\label{eq:family5}
    \fl u_3=\frac{-\mu_1}{2\mu_2}\pm \sqrt{\frac{-6\mu_3}{\mu_2}} \tanh\left(x-\int h(t)dt-\left(\frac{-8\mu_2\mu_3-\mu_1^2}{4\mu_2}\right) t\right).
\end{equation}
\begin{equation}\label{eq:family6}
   \fl u_4=\frac{-\mu_1}{2\mu_2}\pm \sqrt{\frac{-6\mu_3}{\mu_2}} \coth\left(x-\int h(t)dt-\left(\frac{-8\mu_2\mu_3-\mu_1^2}{4\mu_2}\right) t\right).
\end{equation}
\begin{equation}\label{eq:family38}
    \fl u_5=\frac{-\mu_1}{2\mu_2}\pm \sqrt{\frac{-6\mu_3}{\mu_2}} \mathrm{tanh}(x-\int h(t)dt-\alpha t)\pm \sqrt{\frac{-6\mu_3}{\mu_2}} \mathrm{coth}(x-\int h(t)dt-\alpha
    t),
\end{equation}
with
\begin{equation}\label{eq:constraint_38}
    \fl \alpha=\left(\frac{-4\mu_2\mu_3(\pm 6+2)-\mu_1^2}{4\mu_2}\right).
\end{equation}
%==========================================================================================
%==========================================================================================
%=======================           Triangular periodic solutions      ===============
%===========================================================================================
%===========================================================================================
Case 2: Triangular periodic solutions
\begin{equation}\label{eq:family3}
    \fl u_6=\frac{-\mu_1}{2\mu_2}\pm \sqrt{\frac{-6\mu_3}{\mu_2}} \sec\left(x-\int h(t)dt-\left(\frac{-4\mu_2\mu_3-\mu_1^2}{4\mu_2}\right) t\right).
\end{equation}
\begin{equation}\label{eq:family4}
    \fl u_7=\frac{-\mu_1}{2\mu_2}\pm \sqrt{\frac{-6\mu_3}{\mu_2}} \csc\left(x-\int h(t)dt-\left(\frac{-4\mu_2\mu_3-\mu_1^2}{4\mu_2}\right) t\right).
\end{equation}

\begin{equation}\label{eq:family7}
    \fl u_8=\frac{-\mu_1}{2\mu_2}\pm \sqrt{\frac{-6\mu_3}{\mu_2}} \tan\left(x-\int h(t)dt-\left(\frac{8\mu_2\mu_3-\mu_1^2}{4\mu_2}\right) t\right).
\end{equation}
\begin{equation}\label{eq:family8}
    \fl u_9=\frac{-\mu_1}{2\mu_2}\pm \sqrt{\frac{-6\mu_3}{\mu_2}} \cot\left(x-\int h(t)dt-\left(\frac{8\mu_2\mu_3-\mu_1^2}{4\mu_2}\right) t\right).
\end{equation}
\begin{equation}\label{eq:family39}
    \fl u_{10}=\frac{-\mu_1}{2\mu_2}\pm \sqrt{\frac{-6\mu_3}{\mu_2}} \mathrm{tan}(x-\int h(t)dt-\alpha t)\pm \sqrt{\frac{-6\mu_3}{\mu_2}} \mathrm{cot}(x-\int h(t)dt-\alpha
    t),
\end{equation}
with
\begin{equation}\label{eq:constraint_39}
\fl    \alpha=\left(\frac{-4\mu_2\mu_3(\pm
6-2)-\mu_1^2}{4\mu_2}\right).
\end{equation}
%==========================================================================================
%==========================================================================================
%======Jacobi elliptic function solutions and combined Jacobi elliptic function solutions==
%===========================================================================================
%===========================================================================================
Case 3: Jacobi elliptic function solutions and combined Jacobi
elliptic function solutions

\begin{equation}\label{eq:family9}
    \fl u_{11}=\frac{-\mu_1}{2\mu_2}\pm \sqrt{\frac{-6\mu_3}{\mu_2}} \mathrm{ns}\left(x-\int h(t)dt-\left(\frac{-4\mu_2\mu_3(1+k^2)-\mu_1^2}{4\mu_2}\right) t\right).
\end{equation}

\begin{equation}\label{eq:family10}
    \fl u_{12}=\frac{-\mu_1}{2\mu_2}\pm \sqrt{\frac{-6\mu_3}{\mu_2}} \mathrm{dc}\left(x-\int h(t)dt-\left(\frac{-4\mu_2\mu_3(1+k^2)-\mu_1^2}{4\mu_2}\right) t\right).
\end{equation}
\begin{equation}\label{eq:family11}
    \fl u_{13}=\frac{-\mu_1}{2\mu_2}\pm \sqrt{\frac{-6\mu_3}{\mu_2}k'^2} \mathrm{nc}\left(x-\int h(t)dt-\left(\frac{4\mu_2\mu_3(k^2-k'^2)-\mu_1^2}{4\mu_2}\right) t\right).
\end{equation}
\begin{equation}\label{eq:family12}
    \fl u_{14}=\frac{-\mu_1}{2\mu_2}\pm \sqrt{\frac{-6\mu_3}{\mu_2}} \mathrm{cs}\left(x-\int h(t)dt-\left(\frac{4\mu_2\mu_3(1+k'^2)-\mu_1^2}{4\mu_2}\right) t\right).
\end{equation}
\begin{equation}\label{eq:family13}
    \fl u_{15}=\frac{-\mu_1}{2\mu_2}\pm \sqrt{\frac{6\mu_3}{\mu_2}k'^2} \mathrm{nd}\left(x-\int h(t)dt-\left(\frac{4\mu_2\mu_3(1+k'^2)-\mu_1^2}{4\mu_2}\right) t\right).
\end{equation}
\begin{equation}\label{eq:family14}
    \fl u_{16}=\frac{-\mu_1}{2\mu_2}\pm \sqrt{\frac{6\mu_3}{\mu_2}k^2k'^2} \mathrm{sd}\left(x-\int h(t)dt-\left(\frac{4\mu_2\mu_3(1-2k'^2)-\mu_1^2}{4\mu_2}\right) t\right).
\end{equation}
\begin{equation}\label{eq:family15}
    \fl u_{17}=\frac{-\mu_1}{2\mu_2}\pm \sqrt{\frac{-6\mu_3}{\mu_2}\frac{k^2-1}{4}} \frac{1+k\mathrm{sn}\left(x-\int h(t)dt-\left(\frac{4\mu_2\mu_3\frac{1+k^2}{2}-\mu_1^2}{4\mu_2}\right) t\right)}{\mathrm{dn}\left(x-\int h(t)dt-\left(\frac{4\mu_2\mu_3\frac{1+k^2}{2}-\mu_1^2}{4\mu_2}\right) t\right)}.
\end{equation}
\begin{equation}\label{eq:family16}
    \fl u_{18}=\frac{-\mu_1}{2\mu_2}\pm \sqrt{\frac{-6\mu_3}{\mu_2}\frac{1-k^2}{4}} \frac{1+\mathrm{sn}\left(x-\int h(t)dt-\left(\frac{4\mu_2\mu_3\frac{1+k^2}{2}-\mu_1^2}{4\mu_2}\right) t\right)}{\mathrm{cn}\left(x-\int h(t)dt-\left(\frac{4\mu_2\mu_3\frac{1+k^2}{2}-\mu_1^2}{4\mu_2}\right) t\right)}.
\end{equation}
\begin{equation}\label{eq:family17}
    \fl u_{19}=\frac{-\mu_1}{2\mu_2}\pm \sqrt{\frac{-6\mu_3}{\mu_2}\frac{k^2}{4}} \frac{1+\mathrm{dn}\left(x-\int h(t)dt-\left(\frac{-4\mu_2\mu_3\frac{1+k'^2}{2}-\mu_1^2}{4\mu_2}\right) t\right)}{k\mathrm{sn}\left(x-\int h(t)dt-\left(\frac{-4\mu_2\mu_3\frac{1+k'^2}{2}-\mu_1^2}{4\mu_2}\right) t\right)}.
\end{equation}
\begin{equation}\label{eq:family18}
    \fl u_{20}=\frac{-\mu_1}{2\mu_2}\pm \sqrt{\frac{-6\mu_3}{\mu_2}(1-k')^2} \frac{k'+\mathrm{dn}^2\left(x-\int h(t)dt-\left(\frac{-8\mu_2\mu_3(1+k'^2)-\mu_1^2}{4\mu_2}\right) t\right)}{k'-\mathrm{dn}^2\left(x-\int h(t)dt-\left(\frac{4\mu_2\mu_3(1+k'^2)-\mu_1^2}{4\mu_2}\right) t\right)}.
\end{equation}
\begin{equation}\label{eq:family19}
    \fl u_{21}=\frac{-\mu_1}{2\mu_2}\pm \sqrt{\frac{24\mu_3}{\mu_2}k'} \frac{k'\pm \mathrm{dn}^2\left(x-\int h(t)dt-\left(\frac{4\mu_2\mu_3(1+k'^2\pm6k')-\mu_1^2}{4\mu_2}\right) t\right)}{2\sqrt{k'}\mathrm{dn}\left(x-\int h(t)dt-\left(\frac{4\mu_2\mu_3(1+k'^2\pm6k')-\mu_1^2}{4\mu_2}\right) t\right)}.
\end{equation}
\begin{equation}\label{eq:family20}
    \fl u_{22}=\frac{-\mu_1}{2\mu_2} \pm \sqrt{\frac{-6\mu_3}{\mu_2}(1+k')^2} \frac{\mathrm{cn}^2(x-\int h(t)dt- \alpha t)+k'\mathrm{sn}^2(x-\int h(t)dt-\alpha t)}{\mathrm{cn}^2(x-\int h(t)dt-\alpha t)-k'\mathrm{sn}^2(x-\int h(t)dt-\alpha
    t)},
\end{equation}
with
\begin{equation}\label{eq:constraint_20}
    \fl \alpha=\left(\frac{-8\mu_2\mu_3(1+k')^2-\mu_1^2}{4\mu_2}\right).
\end{equation}
\begin{eqnarray}\label{eq:family21}
 \fl   u_{23}=\frac{-\mu_1}{2\mu_2}\pm  \sqrt{\frac{-6\mu_3}{\mu_2}\frac{(1+k)^2}{4}} \nonumber \\\times \sqrt {\frac{2\left(1\pm k\mathrm{sn}\left(x-\int h(t)dt-\left(\frac{-\mu_2\mu_3(1+6k+k^2)-\mu_1^2}{4\mu_2}\right) t\right)\right)}{(1+k)\left(1\pm \mathrm{sn}\left(x-\int h(t)dt-\left(\frac{-\mu_2\mu_3(1+6k+k^2)-\mu_1^2}{4\mu_2}\right) t\right)\right)}}.
\end{eqnarray}
\begin{eqnarray}\label{eq:family22}
 \fl   u_{24}=\frac{-\mu_1}{2\mu_2}\pm  \sqrt{\frac{-6\mu_3}{\mu_2}\frac{1-k^2}{4}} \sqrt {\frac{2}{1-k}} \nonumber \\ \times \frac{\sqrt{(1+k\mathrm{sn}(x-\int h(t)dt-\alpha t))(1+\mathrm{sn}(x-\int h(t)dt-\alpha t))}}{ \mathrm{cn}(x-\int h(t)dt-\alpha
    t)},
\end{eqnarray}
with
\begin{equation}\label{eq:constraint_22}
    \fl \alpha=\left(\frac{-\mu_2\mu_3(1-6k+k^2)-\mu_1^2}{4\mu_2}\right).
\end{equation}

\begin{equation}\label{eq:family24}
\fl    u_{25}=\frac{-\mu_1}{2\mu_2}\pm
\sqrt{\frac{-6\mu_3}{\mu_2}k^2} \mathrm{sn}\left(x-\int
h(t)dt-\left(\frac{-4\mu_2\mu_3(1+k^2)-\mu_1^2}{4\mu_2}\right)
t\right).
\end{equation}
\begin{equation}\label{eq:family25}
 \fl   u_{26}=\frac{-\mu_1}{2\mu_2}\pm \sqrt{\frac{-6\mu_3}{\mu_2}k^2} \mathrm{cd}\left(x-\int h(t)dt-\left(\frac{-4\mu_2\mu_3(1+k^2)-\mu_1^2}{4\mu_2}\right) t\right).
\end{equation}
\begin{equation}\label{eq:family26}
 \fl    u_{27}=\frac{-\mu_1}{2\mu_2}\pm \sqrt{\frac{6\mu_3}{\mu_2}k^2} \mathrm{cn}\left(x-\int h(t)dt-\left(\frac{4\mu_2\mu_3(k^2-k'^2)-\mu_1^2}{4\mu_2}\right) t\right).
\end{equation}
\begin{equation}\label{eq:family27}
\fl    u_{28}=\frac{-\mu_1}{2\mu_2}\pm
\sqrt{\frac{-6\mu_3}{\mu_2}k'^2} \mathrm{sc}\left(x-\int
h(t)dt-\left(\frac{4\mu_2\mu_3(1+k'^2)-\mu_1^2}{4\mu_2}\right)
t\right).
\end{equation}
\begin{equation}\label{eq:family28}
 \fl   u_{29}=\frac{-\mu_1}{2\mu_2}\pm \sqrt{\frac{6\mu_3}{\mu_2}} \mathrm{dn}\left(x-\int h(t)dt-\left(\frac{4\mu_2\mu_3(1+k'^2)-\mu_1^2}{4\mu_2}\right) t\right).
\end{equation}
\begin{equation}\label{eq:family29}
 \fl   u_{30}=\frac{-\mu_1}{2\mu_2}\pm \sqrt{\frac{-6\mu_3}{\mu_2}} \mathrm{ds}\left(x-\int h(t)dt-\left(\frac{4\mu_2\mu_3(1-2k'^2)-\mu_1^2}{4\mu_2}\right) t\right).
\end{equation}
\begin{equation}\label{eq:family30}
 \fl   u_{31}=\frac{-\mu_1}{2\mu_2}\pm \sqrt{\frac{-6\mu_3}{\mu_2}\frac{k^2-1}{4}} \frac{\mathrm{dn}\left(x-\int h(t)dt-\left(\frac{4\mu_2\mu_3 \frac{(1+k^2)}{2}-\mu_1^2}{4\mu_2}\right) t\right)}{1+k\mathrm{sn}\left(x-\int h(t)dt-\left(\frac{-4\mu_2\mu_3\frac{(1+k^2)}{2}-\mu_1^2}{4\mu_2}\right) t\right)}.
\end{equation}
\begin{equation}\label{eq:family31}
 \fl   u_{32}=\frac{-\mu_1}{2\mu_2}\pm \sqrt{\frac{-6\mu_3}{\mu_2}\frac{1-k^2}{4}} \frac{\mathrm{cn}\left(x-\int h(t)dt-\left(\frac{4\mu_2\mu_3 \frac{(1+k^2)}{2}-\mu_1^2}{4\mu_2}\right) t\right)}{1+\mathrm{sn}\left(x-\int h(t)dt-\left(\frac{4\mu_2\mu_3 \frac{(1+k^2)}{2}-\mu_1^2}{4\mu_2}\right) t\right)}.
\end{equation}
\begin{equation}\label{eq:family32}
 \fl   u_{33}=\frac{-\mu_1}{2\mu_2}\pm \sqrt{\frac{-6\mu_3}{\mu_2}\frac{k^2}{4}} \frac{k \mathrm{sn}\left(x-\int h(t)dt-\left(\frac{-4\mu_2\mu_3 \frac{(1+k'^2)}{2}-\mu_1^2}{4\mu_2}\right) t\right)}{1+\mathrm{dn}\left(x-\int h(t)dt-\left(\frac{-4\mu_2\mu_3 \frac{(1+k'^2)}{2}-\mu_1^2}{4\mu_2}\right) t\right)}.
\end{equation}
\begin{equation}\label{eq:family33}
 \fl   u_{34}=\frac{-\mu_1}{2\mu_2}\pm \sqrt{\frac{-6\mu_3}{\mu_2}(1+k')^2} \frac{k'-\mathrm{dn}^2\left(x-\int h(t)dt-\left(\frac{-8\mu_2\mu_3 (1+k'^2)-\mu_1^2}{4\mu_2}\right) t\right)}{k'+\mathrm{dn}^2\left(x-\int h(t)dt-\left(\frac{-8\mu_2\mu_3 (1+k'^2)-\mu_1^2}{4\mu_2}\right) t\right)}.
\end{equation}
\begin{equation}\label{eq:family34}
 \fl   u_{35}=\frac{-\mu_1}{2\mu_2}\pm \sqrt{\frac{\pm 6\mu_3}{\mu_2} (1+k')^2} \frac{2\sqrt{k'}\mathrm{dn}\left(x-\int h(t)dt-\left(\frac{4\mu_2\mu_3 (1+k'^2\pm 6k')-\mu_1^2}{4\mu_2}\right) t\right)}{k'\pm \mathrm{dn}^2\left(x-\int h(t)dt-\left(\frac{4\mu_2\mu_3 (1+k'^2\pm 6k')-\mu_1^2}{4\mu_2}\right) t\right)}.
\end{equation}
\begin{equation}\label{eq:family35}
 \fl   u_{36}=\frac{-\mu_1}{2\mu_2}\pm \sqrt{\frac{-6\mu_3}{\mu_2}(1-k')^2} \frac{\mathrm{cn}^2(x-\int h(t)dt-\alpha t)-k'\mathrm{sn}^2(x-\int h(t)dt-\alpha t)}{\mathrm{cn}^2(x-\int h(t)dt-\alpha t)+k'\mathrm{sn}^2(x-\int h(t)dt-\alpha
    t)},
\end{equation}
with
\begin{equation}\label{eq:constraint_35}
  \fl  \alpha=\left(\frac{-8\mu_2\mu_3(1+k')^2-\mu_1^2}{4\mu_2}\right).
\end{equation}
\begin{equation}\label{eq:family36}
 \fl   u_{37}=\frac{-\mu_1}{2\mu_2}\pm \sqrt{\frac{-6\mu_3}{\mu_2}k} \sqrt {\frac{(1+k)\left(1\pm \mathrm{sn}\left(x-\int h(t)dt-\left(\frac{-\mu_2\mu_3 (1+6k+k^2 )-\mu_1^2}{4\mu_2}\right) t\right)\right)}{2\left(1\pm k\mathrm{sn}\left(x-\int h(t)dt-\left(\frac{-\mu_2\mu_3 (1+6k+k^2 )-\mu_1^2}{4\mu_2}\right) t\right)\right)}}.
\end{equation}
\begin{eqnarray}\label{eq:family37}
  \fl  u_{38}=\frac{-\mu_1}{2\mu_2}\pm  \sqrt{\frac{-6\mu_3}{\mu_2}(1-k)^2} \sqrt {\frac{1-k}{2}} \nonumber \\ \times \frac{\mathrm{cn}(x-\int h(t)dt-\alpha t)}{\sqrt{(1+k\mathrm{sn}(x-\int h(t)dt-\alpha t))(1+\mathrm{sn}(x-\int h(t)dt-\alpha
    t))}},
\end{eqnarray}
with
\begin{equation}\label{eq:constraint_37}
   \fl \alpha=\left(\frac{-\mu_2\mu_3(1-6k+k^2 )-\mu_1^2}{4\mu_2}\right).
\end{equation}

\begin{equation}\label{eq:family40}
\fl    u_{39}=\frac{-\mu_1}{2\mu_2}\pm \sqrt{\frac{-6\mu_3}{\mu_2}}
\mathrm{ns}(x-\int h(t)dt-\alpha t)\pm
\sqrt{\frac{-6\mu_3}{\mu_2}k^2} \mathrm{sn}(x-\int h(t)dt-\alpha
    t),
\end{equation}
with
\begin{equation}\label{eq:constraint_40}
 \fl   \alpha=\left(\frac{-4\mu_2\mu_3(\pm 6\sqrt{k^2}+1+k^2)-\mu_1^2}{4\mu_2}\right).
\end{equation}
\begin{equation}\label{eq:family41}
\fl    u_{40}=\frac{-\mu_1}{2\mu_2}\pm \sqrt{\frac{-6\mu_3}{\mu_2}}
\mathrm{dc}(x-\int h(t)dt-\alpha t)\pm
\sqrt{\frac{-6\mu_3}{\mu_2}k^2} \mathrm{cd}(x-\int h(t)dt-\alpha
    t),
\end{equation}
with
\begin{equation}\label{eq:constraint_41}
 \fl   \alpha=\left(\frac{-4\mu_2\mu_3(\pm 6\sqrt{k^2}+1+k^2)-\mu_1^2}{4\mu_2}\right).
\end{equation}
\begin{equation}\label{eq:family42}
\fl    u_{41}=\frac{-\mu_1}{2\mu_2}\pm
\sqrt{\frac{-6\mu_3}{\mu_2}k'^2} \mathrm{nc}(x-\int h(t)dt-\alpha
t)\pm \sqrt{\frac{6\mu_3}{\mu_2}k^2} \mathrm{cn}(x-\int
h(t)dt-\alpha
    t),
\end{equation}
with
\begin{equation}\label{eq:constraint_42}
 \fl   \alpha=\left(\frac{-4\mu_2\mu_3(\pm 6\sqrt{-k^2k'^2}+k'^2-k^2)-\mu_1^2}{4\mu_2}\right).
\end{equation}
\begin{equation}\label{eq:family43}
\fl    u_{42}=\frac{-\mu_1}{2\mu_2}\pm \sqrt{\frac{-6\mu_3}{\mu_2}}
\mathrm{cs}(x-\int h(t)dt-\alpha t)\pm
\sqrt{\frac{-6\mu_3}{\mu_2}k'^2} \mathrm{sc}(x-\int h(t)dt-\alpha
    t),
\end{equation}
with
\begin{equation}\label{eq:constraint_43}
  \fl  \alpha=\left(\frac{-4\mu_2\mu_3(\pm 6\sqrt{k'^2}-1-k'^2)-\mu_1^2}{4\mu_2}\right).
\end{equation}
\begin{equation}\label{eq:family44}
\fl    u_{43}=\frac{-\mu_1}{2\mu_2}\pm
\sqrt{\frac{6\mu_3}{\mu_2}k'^2} \mathrm{nd}(x-\int h(t)dt-\alpha
t)\pm \sqrt{\frac{6\mu_3}{\mu_2}} \mathrm{dn}(x-\int h(t)dt-\alpha
    t),
\end{equation}
with
\begin{equation}\label{eq:constraint_44}
 \fl   \alpha=\left(\frac{-4\mu_2\mu_3(\pm 6\sqrt{k'^2}-1-k'^2)-\mu_1^2}{4\mu_2}\right).
\end{equation}
\begin{equation}\label{eq:family45}
\fl    u_{44}=\frac{-\mu_1}{2\mu_2}\pm
\sqrt{\frac{6\mu_3}{\mu_2}k^2k'^2} \mathrm{sd}(x-\int h(t)dt-\alpha
t)\pm \sqrt{\frac{-6\mu_3}{\mu_2}} \mathrm{ds}(x-\int h(t)dt-\alpha
    t),
\end{equation}
with
\begin{equation}\label{eq:constraint_45}
 \fl   \alpha=\left(\frac{4\mu_2\mu_3(\pm 6\sqrt{-k^2k'^2}-1+2k'^2)-\mu_1^2}{4\mu_2}\right).
\end{equation}
\begin{eqnarray}\label{eq:family46}
  \fl  u_{45}=\frac{-\mu_1}{2\mu_2}\pm \sqrt{\frac{-6\mu_3}{\mu_2}\frac{k^2-1}{4}} \frac{1+k\mathrm{sn}(x-\int h(t)dt-\alpha t)}{\mathrm{dn}(x-\int h(t)dt-\alpha t)} \nonumber \\ \pm   \sqrt{\frac{-6\mu_3}{\mu_2}\frac{k^2-1}{4}} \frac{\mathrm{dn}(x-\int h(t)dt-\alpha t)}{1+k\mathrm{sn}(x-\int h(t)dt-\alpha
    t)},
\end{eqnarray}
with
\begin{equation}\label{eq:constraint_46}
  \fl   \alpha=\left(\frac{-4\mu_2\mu_3(\pm \frac{3}{2}-\frac{1}{2})(k^2-1)-\mu_1^2}{4\mu_2}\right).
\end{equation}
\begin{eqnarray}\label{eq:family47}
\fl    u_{46}=\frac{-\mu_1}{2\mu_2}\pm
\sqrt{\frac{-6\mu_3}{\mu_2}\frac{1-k^2}{4}}
\frac{1+\mathrm{sn}(x-\int h(t)dt-\alpha t)}{\mathrm{cn}(x-\int
h(t)dt-\alpha t)} \nonumber \\ \pm
\sqrt{\frac{-6\mu_3}{\mu_2}\frac{1-k^2}{4}} \frac{\mathrm{cn}(x-\int
h(t)dt-\alpha t)}{1+\mathrm{sn}(x-\int h(t)dt-\alpha
    t)},
\end{eqnarray}
with
\begin{equation}\label{eq:constraint_47}
 \fl   \alpha=\left(\frac{-4\mu_2\mu_3\left(\frac{3}{2}(1-k^2)-\frac{1}{2}(1+k^2)\right)-\mu_1^2}{4\mu_2}\right).
\end{equation}
\begin{eqnarray}\label{eq:family48}
\fl    u_{47}=\frac{-\mu_1}{2\mu_2}\pm
\sqrt{\frac{-6\mu_3}{\mu_2}\frac{k^2}{4}} \frac{1+\mathrm{dn}(x-\int
h(t)dt-\alpha t)}{k \mathrm{sn}(x-\int h(t)dt-\alpha t)} \nonumber
\\ \pm \sqrt{\frac{-6\mu_3}{\mu_2}\frac{k^2}{4}} \frac{k
\mathrm{sn}(x-\int h(t)dt-\alpha t)}{1+\mathrm{dn}(x-\int
h(t)dt-\alpha
    t)},
\end{eqnarray}
with
\begin{equation}\label{eq:constraint_48}
 \fl    \alpha=\left(\frac{-\mu_2\mu_3(\pm 3k^2+2(1+k'^2))-\mu_1^2}{4\mu_2}\right).
\end{equation}
\begin{eqnarray}\label{eq:family49}
\fl    u_{48}=\frac{-\mu_1}{2\mu_2}\pm
\sqrt{\frac{-6\mu_3}{\mu_2}(1-k')^2} \frac{k'+\mathrm{dn}^2(x-\int
h(t)dt-\alpha t)}{k'-\mathrm{dn}^2(x-\int h(t)dt-\alpha t)}
\nonumber \\ \pm \sqrt{\frac{-6\mu_3}{\mu_2}(1+k')2}
\frac{k'-\mathrm{dn}^2(x-\int h(t)dt-\alpha
t)}{k'+\mathrm{dn}^2(x-\int h(t)dt-\alpha
    t)},
\end{eqnarray}
with
\begin{equation}\label{eq:constraint_49}
\fl \alpha=\left(\frac{-4\mu_2\mu_3(\pm 6\sqrt{(1-k')^2(1+k'^2)}
+2(1+k'^2))-\mu_1^2}{4\mu_2}\right).
\end{equation}
\begin{eqnarray}\label{eq:family50}
\fl    u_{49}=\frac{-\mu_1}{2\mu_2}\pm
\sqrt{\frac{24\mu_3}{\mu_2}k'} \frac{k'\pm \mathrm{dn}^2(x-\int
h(t)dt-\alpha t)}{2\sqrt{k'}\mathrm{dn}(x-\int h(t)dt-\alpha t)}
\nonumber \\ \pm \sqrt{\frac{-6\mu_3}{\mu_2}(\mp (1+k')^2)}
\frac{2\sqrt{k'}\mathrm{dn}(x-\int h(t)dt-\alpha t)}{k'\pm
\mathrm{dn}^2(x-\int h(t)dt-\alpha
    t)},
\end{eqnarray}
with
\begin{equation}\label{eq:constraint_50}
 \fl   \alpha=\left(\frac{-4\mu_2\mu_3(\pm 6\sqrt{\pm8k'(1+k')^2}-(1+k'^2\pm6k'))-\mu_1^2}{4\mu_2}\right).
\end{equation}
\begin{eqnarray}\label{eq:family51}
\fl    u_{50}=\frac{-\mu_1}{2\mu_2}\pm
\sqrt{\frac{-6\mu_3}{\mu_2}(1+k')^2} \frac{\mathrm{cn}^2(x-\int
h(t)dt-\alpha t)+k'\mathrm{sn}^2(x-\int h(t)dt-\alpha
t)}{\mathrm{cn}^2(x-\int h(t)dt-\alpha t)-k'\mathrm{sn}^2(x-\int
h(t)dt-\alpha t)}  \nonumber \\ \pm
\sqrt{\frac{-6\mu_3}{\mu_2}(1-k')^2} \frac{\mathrm{cn}^2(x-\int
h(t)dt-\alpha t)-k'\mathrm{sn}^2(x-\int h(t)dt-\alpha
t)}{\mathrm{cn}^2(x-\int h(t)dt-\alpha t)+k'\mathrm{sn}^2(x-\int
h(t)dt-\alpha t)},
\end{eqnarray}
with
\begin{equation}\label{eq:constraint_51}
 \fl   \alpha=\left(\frac{-4\mu_2\mu_3(\pm 6(1-k'^2)+2(1+k')^2)-\mu_1^2}{4\mu_2}\right).
\end{equation}
\begin{eqnarray}\label{eq:family52}
\fl    u_{51}=\frac{-\mu_1}{2\mu_2}\pm
\sqrt{\frac{-6\mu_3}{\mu_2}\frac{(1+k)^2}{4}} \sqrt {\frac{2(1\pm
k\mathrm{sn}(x-\int h(t)dt-\alpha t))}{(1+k)((1\pm
\mathrm{sn}(x-\int h(t)dt-\alpha t)))}} \nonumber \\ \pm
\sqrt{\frac{-6\mu_3}{\mu_2}k} \sqrt {\frac{(1+k)((1\pm
\mathrm{sn}(x-\int h(t)dt-\alpha t)))}{2(1\pm k \mathrm{sn}(x-\int
h(t)dt-\alpha
    t))}},
\end{eqnarray}
with
\begin{equation}\label{eq:constraint_52}
\fl \alpha=\left(\frac{-\mu_2\mu_3(\pm
12\sqrt{k(1+k)^2}+1+6k+k^2)-\mu_1^2}{4\mu_2}\right).
\end{equation}
\begin{eqnarray}\label{eq:family53}
\fl    u_{52}=\frac{-\mu_1}{2\mu_2}\pm
\sqrt{\frac{-6\mu_3}{\mu_2}\frac{1-k^2}{4}} \sqrt {\frac{2}{1-k}}\pm
\sqrt{\frac{-6\mu_3}{\mu_2}(1-k)^2}\nonumber \\ \fl \times \sqrt
\frac{\sqrt{(1+k\mathrm{sn}(x-\int h(t)dt-\alpha
t))(1+\mathrm{sn}(x-\int h(t)dt-\alpha t))}}{ \mathrm{cn}(x-\int
h(t)dt-\alpha t)} \nonumber \\ \fl \pm
\sqrt{\frac{-6\mu_3}{\mu_2}(1-k)^2} \sqrt
{\frac{1-k}{2}}\frac{\mathrm{cn}(x-\int h(t)dt-\alpha
t)}{\sqrt{(1+k\mathrm{sn}(x-\int h(t)dt-\alpha
t))(1+\mathrm{sn}(x-\int h(t)dt-\alpha
    t))}},
\end{eqnarray}
with
\begin{equation}\label{eq:constraint_53}
 \fl   \alpha=\left(\frac{-\mu_2\mu_3(\pm 6\sqrt{2(1-k^2)(1-k)^2}+1-6k+k^2)-\mu_1^2}{4\mu_2}\right).
\end{equation}
%======================================================================================
%======================================================================================
%=======================rational solution=============================================
%======================================================================================
%======================================================================================
Case 4: rational solution
\begin{equation}\label{eq:family23}
\fl    u_{53}=\frac{-\mu_1}{2\mu_2}\pm \sqrt{\frac{-6\mu_3}{\mu_2}}
\frac {1}{x-\int h(t)dt-\left(\frac{-\mu_1^2}{4\mu_2}\right) t}.
\end{equation}
All the above solutions can be considered only in the case when
$\mu_2\ne 0$.
%======================================================================================
%======================================================================================
%=======================Applications=============================================
%======================================================================================
%======================================================================================
\section{Applications to arterial mechanics}\label{subsec:Applications}
\begin{figure}
\centering
  \includegraphics*[width=0.5\linewidth]{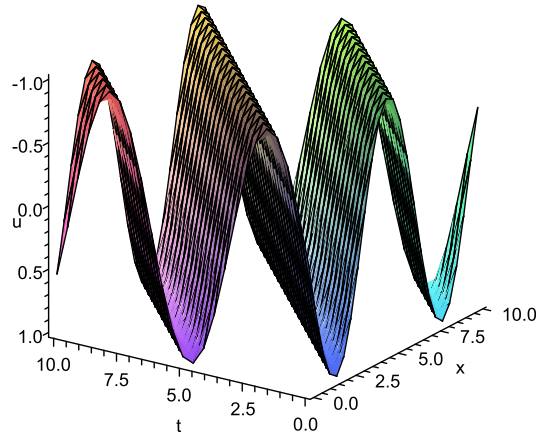}
  \caption{Periodic wave solution for the mKdV equation which is given by family24 (with $\mu_2=1$, $\mu_3=-1$, $h(t)=1$ and $k=0.5$. }\label{fig:periodic}
\end{figure}

\begin{figure}
\centering
\includegraphics*[width=0.5\linewidth]{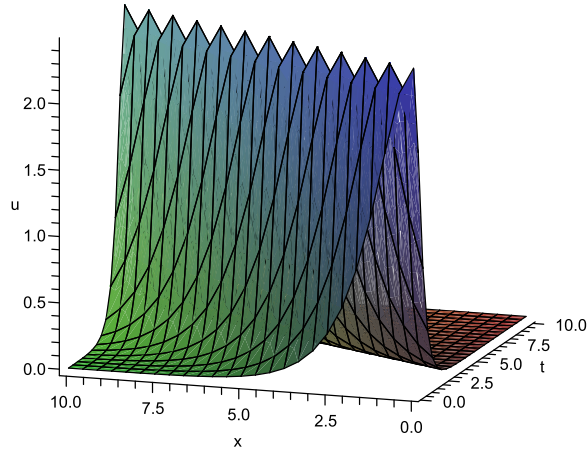}\\
  \caption{ Solitary wave solution for the mKdV equation which is given by family2 (with $\mu_2=1$, $\mu_3=1$ and $h(t)=1$.  }\label{fig:solitary}
\end{figure}

\begin{figure}
\centering
\includegraphics*[width=0.5\linewidth]{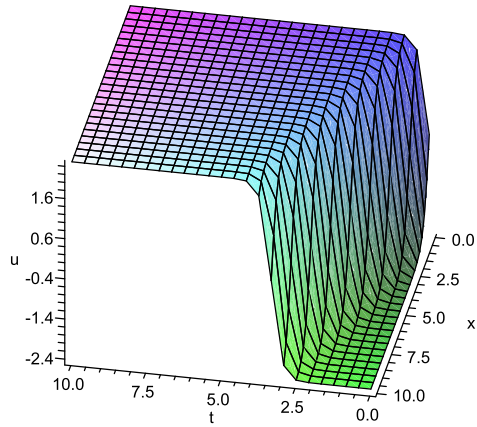}\\
  \caption{Anti-kink wave solution for the mKdV equation which is given by family5 (with $\mu_2=1$, $\mu_3=-1$ and $h(t)=1$. }\label{fig:kink}
\end{figure}

In arterial mechanics \cite{Demiray20092}, treating the arteries as
a thin walled prestressed elastic tube with variable radius (or,
with stenosis in \cite{Demiray20071}) and blood as an inviscid
fluid, the governing equation which models the weakly nonlinear
waves in such a fluid-filled elastic tubes is the
variable-coefficient mKdV equation \eref{eq:vcmKDV}. Also, in \cite
{Demiray20041,kudryashov2008,Demiray20051} treating the arteries as
a thin walled prestressed elastic tube and blood as an
incompressible inviscid fluid , the governing equation which models
the weakly nonlinear waves in such a fluid-filled elastic tubes is
the mKdV equation \eref{eq:mKDV}. We obtained some useful solutions
for these two equations such as periodic solution given by
\eref{eq:family24} and solitary wave solution given by
\eref{eq:family2} and anti-kink wave solution given by \eref{eq:family5} which are shown in, (with $h(t)=\mathrm{const.}$), fig.\ref{fig:periodic}, fig.\ref{fig:solitary} and fig.\ref{fig:kink} respectively.\\

%======================================================================================
%======================================================================================
%=======================Concluding remarks=============================================
%======================================================================================
%======================================================================================
\section{Concluding remarks}\label{subsec:concluding remarks}

 Using the classical symmetry analysis method we obtained two similarity transformations \eref{eq:simtransform1} and \eref{eq:simtransform} that, for arbitrary $h(t)$,
 transforms the original nonlinear PDE \eref{eq:MKDV} into two nonlinear ODEs
 \eref{eq:ode1} and \eref{eq:ode} respectively. Then we used the generalized Jacobi elliptic function expansion
method to obtain many solutions for \eref{eq:ode}. The set of
solutions from $u_1$ to $u_4$, from $u_6$ to $u_9$, from $u_{11}$ to
$u_{38}$ and $u_{53}$ with $h(t)=0$, can be found in
\cite{Malfliet2004,Fu2004,Wazwaz2007,Liu2009,bekir2009}. To our best
knowledge, the set of solutions $u_5$, $u_{10}$ and from $u_{39}$ to
$u_{52}$ are new solutions of \eref{eq:MKDV} and are not shown in
the current literature until now. \\ The generalized
variable-coefficient Gardner equation
\begin{equation}\label{eq:Ggardner}
    u_t+a(t) u u_x+b(t)u^2 u_x+h_1(t)u_{xxx}+d(t)u_x+f(t)u=0
\end{equation}
was considered in \cite{Zhang2008} and solved with a new generalized
algebraic method. Also, the set of solutions from $u_1$ to $u_4$,
from $u_6$ to $u_9$, from $u_{11}$ to $u_{38}$ and $u_{53}$ (with
$a(t)=\mu1$, $b(t)=\mu2$, $h_1(t)=\mu3$,$d(t)=h(t)$ and $f(t)=0$),
are obtained in \cite{Zhang2008}. But, the set of solutions $u_5$,
$u_{10}$ and from $u_{39}$ to
$u_{52}$ are not obtained in \cite{Zhang2008}. \\

\appendix
\section{Some special solutions of Equation \eref{eq:ellipticeq}.}\label{S:solode}
\begin{tabular}{c c c c} \hline
$r$&$p$&$q$&$\phi$ \\ \hline
0&1&0&$e^\pm\zeta$\\
1&1&0&$\mathrm{sinh}(\zeta)$\\
-1&1&0&$\mathrm{cosh}(\zeta)$\\
1&-1&0&$\mathrm{sin}(\zeta)$,$\mathrm{cos}(\zeta)$\\
0&0&1&$\frac{1}{\zeta}$\\
0&1&-1&$\mathrm{sech}(\zeta)$\\
0&1&1&$\mathrm{csch}(\zeta)$\\
1&-2&1&$\mathrm{tanh}(\zeta)$,$\mathrm{coth}(\zeta)$\\
1&2&1&$\mathrm{tan}(\zeta)$,$\mathrm{cot}(\zeta)$\\
0&-1&1&$\mathrm{sec}(\zeta)$,$\mathrm{csc}(\zeta)$\\
1&$-(1+k^2)$&$k^2$&$\mathrm{sn}(\zeta)$\\
$k'^2$&$k^2-k'^2$&$-k^2$&$\mathrm{cn}(\zeta)$\\
1&$1+k'^2$&$k'^2$&$sc(\zeta)=\frac{\mathrm{sn}(\zeta)}{\mathrm{cn}(\zeta)}$\\
$-k'^2$&$1+k'^2$&-1&$\mathrm{dn}(\zeta)$\\
$-k^2k'^2$&$1-2k'^2$&1&$\frac{\mathrm{dn}(\zeta)}{\mathrm{sn}(\zeta)}$\\
$\frac{k^2-1}{4}$&$\frac{1+k^2}{2}$&$\frac{k^2-1}{4}$&$\frac{\mathrm{dn}(\zeta)}{(1+k\mathrm{sn}(\zeta))}$\\
$\frac{1-k^2}{4}$&$\frac{1+k^2}{2}$&$\frac{1-k^2}{4}$&$\frac{\mathrm{cn}(\zeta)}{(1+\mathrm{sn}(\zeta))}$\\
$\frac{k^2}{4}$&$-\frac{1+k'^2}{2}$&$\frac{k^2}{4}$&$\frac{k\mathrm{sn}(\zeta)}{(1+\mathrm{dn}(\zeta))}$\\
$(1-k')^2$&$-2(1+k'^2)$&$(1+k'^2)$&$\frac{k'-\mathrm{dn}^2(\zeta)}{(k'+\mathrm{dn}^2(\zeta))}$\\
$-4k'$&$(1+k^2\pm 6k')$&$\mp(1+k')^2$&$\frac{2\sqrt{k'}\mathrm{dn}(\zeta)}{(k'\pm \mathrm{dn}^2(\zeta))}$\\
$(1+k')^2$&$-2(1+k')^2$&$(1-k')^2$&$\frac{\mathrm{cn}^2(\zeta)-k'\mathrm{sn}^2(\zeta)}{\mathrm{cn}^2(\zeta)+k'\mathrm{sn}^2(\zeta)}$\\
$\frac{(1+k)^2}{4}$&$-\frac{1+6k+k^2}{4}$&$k$&$\sqrt{\frac {(1+k)(1\pm \mathrm{sn}(\zeta))}{2(1\pm k\mathrm{sn}(\zeta))}}$\\
$\frac{1-k^2}{4}$&$-\frac{1-6k+k^2}{4}$&$\frac{(1-k)^2}{2}$&$\sqrt{\frac{1-k}{2}}\frac{\mathrm{cn}(\zeta)}{\sqrt{(1+k\mathrm{sn}(\zeta))(1+\mathrm{sn}(\zeta))}}$\\
\hline
\\
where $k^2+k'^2=1.$
\end{tabular}

%\section*{Acknowledgments}
\ack {I wish to express my thanks, respectively, to Prof. Yu. Yu.
Tarasevich, Prof. A. I. Lobanov and Prof. A. G. Kushner for valuable
discussions concerning this work.}
\\

\end{document}